\documentclass[11pt]{article}
\usepackage{epsf}

\def\qb{{\bar{q}}}

\def\e{\epsilon}

\def\eps{\epsilon}

\def\Ord{{\cal O}}
\def\cm{{\cal M}}
\def\cmt{\tilde{\cal M}}
\def\Ft{\tilde{F}}
\def\Re{{\rm Re}}
\newcommand{\la}{\langle}
\newcommand{\ra}{\rangle}
\def\RS{{\scriptscriptstyle\rm R\!.S\!.}}
\def\CDR{{\scriptscriptstyle\rm CDR}}
\def\bom#1{{\mbox{\boldmath $#1$}}}
\def\MSbar{\overline{\rm MS}}
\def\Boxsixint{{\rm Box}^{(6)}}
\def\Triint{{\rm Tri}}
\def\lr{\leftrightarrow}
\def\li#1{{\mathop{\rm Li}\nolimits}_#1}
\def\Li{\mathop{\rm Li}\nolimits}
\def\fig#1{fig.~{\ref{#1}}}
\def\eqn#1{eq.~(\ref{#1})}

\newskip\humongous \humongous=0pt plus 1000pt minus 1000pt
\def\caja{\mathsurround=0pt}
\def\eqalign#1{\,\vcenter{\openup1\jot \caja
        \ialign{\strut \hfil$\displaystyle{##}$&$
        \displaystyle{{}##}$\hfil\crcr#1\crcr}}\,}
\newif\ifdtup

\newcounter{eqnumber}[section]
\renewcommand{\theeqnumber}{\thesection.\arabic{eqnumber}}
\def\equn{
\refstepcounter{eqnumber}
\eqno({\rm \theeqnumber})
}

\begin{document}

\begin{titlepage}

\begin{flushright}
  hep-ph/0010075 \hfill     SLAC--PUB--8655\\
          UCLA/00/TEP/26\\
          October, 2000
\end{flushright}

\begin{center}
\begin{Large}
{\bf Two-Loop Correction to Bhabha Scattering} 
\end{Large}

\vskip 1.5cm

{Z. Bern$^\star$\\
\it Department of Physics and Astronomy \\
UCLA, Los Angeles, CA 90095-1547}

\vskip 0.7cm

{L. Dixon$^\dagger$\\
\it Stanford Linear Accelerator Center\\
Stanford University\\
Stanford, CA 94309}

\vskip 0.6cm
and 
\vskip 0.4cm

{A. Ghinculov$^\star$\\
\it Department of Physics and Astronomy \\
UCLA, Los Angeles, CA 90095-1547}

\end{center}

\vskip 2 cm 
\begin{abstract}
We present the two-loop virtual QED corrections to $e^+e^- \to
\mu^+\mu^-$ and Bhabha scattering in dimensional regularization.  The
results are expressed in terms of polylogarithms.  The form of the
infrared divergences agrees with previous expectations.  These results
are a crucial ingredient in the complete next-to-next-to-leading order
QED corrections to these processes. A future application will be to
reduce theoretical uncertainties associated with luminosity
measurements at $e^+ e^-$ colliders.  The calculation also tests
methods that may be applied to analogous QCD processes.
\end{abstract}


\vskip 1cm
\begin{center}
{\sl Submitted to Physical Review D}
\end{center}

\vfill
\noindent\hrule width 3.6in\hfil\break
\begin{small}
${}^{\star}$Research supported by the US Department of Energy under grant 
DE-FG03-91ER40662.\hfil\break
${}^{\dagger}$Research supported by the US Department of Energy under grant 
DE-AC03-76SF00515.\hfil\break
\end{small}

\end{titlepage}
             
\baselineskip 16pt


\renewcommand{\thefootnote}{\arabic{footnote}}
\setcounter{footnote}{0}


\section{Introduction}
\label{IntroSection}

Bhabha scattering is an important process for extracting physics from
experiments at electron-positron colliders primarily because it provides an
effective means for determining luminosity.  These measurements depend
on having precise theoretical predictions for the Bhabha scattering
cross sections.  As yet, the complete next-to-next-to-leading order
(NNLO) QED corrections needed for reducing theoretical uncertainties
have not been computed.  In this paper we present the complete
two-loop matrix elements that would enter into such a computation.
This calculation also provides a means for validating techniques that
can be applied to physically important but more intricate QCD
calculations.  It also provides an additional explicit verification of
a general formula due to Catani~\cite{Catani} for the structure of 
two-loop infrared divergences, and allows us to determine 
the process-dependent terms for the processes at hand.

In Bhabha scattering there are two distinct kinematic
regions: small angle Bhabha scattering (SABS), and large angle (LABS).
In the LEP/SLC energy range, SABS is used to measure the machine
luminosity via a dedicated small angle luminosity detector.  SABS
has a large cross section --- about four times larger than $Z$ decay 
in the $1^\circ-3^\circ$ window ---
making it particularly effective as a luminosity monitor.
At the same time, SABS is calculable theoretically with high accuracy
from known physics (mainly QED), apart from hadronic vacuum polarization
corrections that rely upon the experimental data for $e^+e^-$
annihilation into hadrons at low energy~\cite{AFKLMT,SABSRef}.
Therefore, SABS is an 
important ingredient in measuring any absolute cross section. 
For instance, the measurement of the hadronic cross section at the $Z$
peak, $\sigma_{\rm h}^o$, which enters several precision observables,
is especially dependent on an accurate theoretical understanding of 
Bhabha scattering.

At LEP/SLC, large angle Bhabha scattering interferes with $e^+e^- \to Z
\to e^+e^-$ and so it is needed to disentangle important parameters such
as the electroweak mixing angle. It is also useful for measuring the
luminosity at flavor factories such as BABAR, BELLE, DA$\Phi$NE, VEPP-2M,
and BEPC/BES~\cite{MesonFact}.  A peculiarity of future electron linear
colliders is that the luminosity spectrum is not monochromatic because of
the beam-beam effect. Because of this, measuring the total small angle
cross section of Bhabha scattering alone is not sufficient, and therefore
the angular distribution of LABS was proposed for disentangling the
luminosity spectrum~\cite{Spectrum}.

Due to the experimental importance of this process, significant effort
has been devoted to developing Monte Carlo event generators --- see
for instance ref. \cite{mceg} for an overview.  In order to match the
impressive experimental precision, a complete inclusion of NNLO QED
quantum effects has become necessary. On the theoretical side,
however, the calculation of two-loop four-point amplitudes has been a
roadblock to further progress.

In this article we present the two-loop virtual QED corrections to
the differential cross section for Bhabha scattering, i.e., 
the two-loop amplitude interfered with the tree amplitude and summed 
over all spins.  We neglect the small
electron mass in comparison to all other kinematic invariants, and use
dimensional regularization to handle the ensuing infrared
divergences.  Besides these contributions, a number of other virtual
and real emission contributions (discussed in the conclusions) still 
need to be obtained before a full Monte Carlo program for the
Bhabha scattering cross section can be constructed.

The two-loop QED four-fermion amplitudes are also a useful testing ground
for two-loop QCD calculations containing more than one kinematic
invariant, which are required for higher-order jet cross sections
and other aspects of collider physics.  For processes that depend on
a single momentum invariant, a number of important quantities have
been calculated up to four loops, such as the total cross section for
$e^+e^-$ annihilation into hadrons and the 
QCD $\beta$-function~\cite{FourLoopPrevious}.  In contrast,
the only complete two-loop four-point scattering amplitudes presently
known for generic kinematics in massless gauge theory are the $N=4$
super-Yang-Mills amplitudes~\cite{BRY,BDDPR}, and  $gg \to gg$
in a single helicity configuration in pure gauge theory~\cite{AllplusTwo}.
The two-loop amplitudes required for NNLO computations of jet production 
in hadron colliders, or for NNLO three-jet rates and other event shape
variables at $e^+e^-$ colliders, remain uncalculated.  We note in
passing that partial results for the leading-color part of two-loop
contributions to 
quark-quark scattering have very recently appeared~\cite{GloverTY}.

Two important technical breakthroughs are the calculations
of the dimensionally regularized scalar double box integrals with 
planar~\cite{PBScalar} and non-planar~\cite{NPBScalar} topologies 
and all external legs massless, and the development of reduction 
algorithms for the same types of integrals with loop momenta in the 
numerator (tensor integrals)~\cite{PBReduction,NPBReduction,GRReduction,
ReductionTalks,GloverTY}.  Related integrals, which also arise in the 
reduction procedure, have been computed in refs.~\cite{AGOOne,AGOTwo}.
Taken together, these results are sufficient to compute all loop integrals 
required for $2 \rightarrow 2$ massless scattering amplitudes at two loops,
thus removing a major obstacle to several types of NNLO calculations.  
In this paper we use these techniques to evaluate the integrals encountered
in the Bhabha calculation.  An even more recent result concerning two-loop
planar double box integrals with one massive external
leg~\cite{TwoloopOneMassIntegrals} holds promise for the NNLO computation
of three-jet rates at $e^+e^-$ colliders.

There has also been significant progress in developing general formalisms
for other aspects of NNLO computations involving massless particles.  
The motivation has typically been infrared-safe observables in QCD, 
but many of the developments can be applied to the Bhabha process as well.
The developments include an understanding of the intricate structure of the 
infrared singularities that arise when more than one particle is 
unresolved (i.e., is soft or collinear with another 
particle)~\cite{LoopSplit,ThreeSplit,GDRGlover}.
Improved approximations to the NNLO correction to splitting functions 
have been constructed recently as well~\cite{APSplitting}.  

Infrared divergences are a significant complication in all the 
QCD and QED computations mentioned above.  In any suitably 
``infrared-safe'' observable all final-state divergences
will cancel~\cite{LeeNauenberg}.  However, divergences occur in
individual amplitudes for fixed particle number, and it is very useful
to have a general description of such divergences. 
Catani has presented a general formula for the infrared divergence 
appearing in any two-loop QCD amplitude~\cite{Catani}.
By appropriately adjusting group theory factors, it is straightforward
to convert Catani's QCD formula to a QED formula, allowing us to
directly verify it.  Moreover, we extract the exact form of
a process-dependent term in the formula, for the case of QED scattering
of four charged fermions.  Previously, the only process for which
this term had been extracted~\cite{Catani} was the quark form factor 
which enters Drell-Yan production~\cite{QuarkFormFactor}.  (It should
also now be possible to extract it for $gg \to$~Higgs using the recent 
two-loop computation~\cite{HarlanderggH}.)
Interestingly, a simple generalization of the quark form factor term
(converted to QED) correctly predicts the process-dependent term for the
$e^+e^- \to \mu^+\mu^-$ and Bhabha amplitudes.
We also use Catani's formula to conveniently organize the infrared 
divergences and to absorb some of the finite terms.

The previously computed non-abelian gauge theory
amplitudes~\cite{BRY,BDDPR,AllplusTwo} were obtained via cutting
methods.  The low multiplicity and relative simplicity 
of the $e^+e^- \to \mu^+\mu^-$ and Bhabha scattering Feynman diagrams 
makes it relatively easy to directly compute the diagrams, as we do here.
We include here only the pure QED diagrams, neglecting for example the
contributions of $Z$ exchange, and hadronic vacuum polarization effects.
The former are negligible at this order in SABS and in LABS at flavor 
factories.  The hadronic contributions are important, but much of their 
effect is straightforward to include by introducing a running coupling.

We perform the calculation in dimensional regularization~\cite{HV}
with $d=4-2\e$ and set the small electron mass to zero, since it 
is the only form in which the required two-loop momentum integrals are known. 
Moreover, it provides a powerful method for simultaneously dealing
with both the infrared and ultraviolet divergences encountered in
gauge theories.  Traditionally, dimensional regularization is not used
for QED, in part because the infrared divergences are relatively tame
compared to non-abelian gauge theories, so photon and electron masses 
are sufficient for cutting off the theory.  Another important reason for 
using dimensional regularization is to validate techniques that can 
also be applied to the more complicated case of QCD.  In QCD, dimensional 
regularization is the universally utilized method for dealing with 
divergences.  

In the high-energy Bhabha process, even with an
``infrared-safe'' (calorimetric) final-state definition, the 
electron mass will still appear in large logarithms of the form 
$L \equiv \ln(Q^2/m_e^2)$ due to initial-state radiation.
However, in the dimensionally regulated amplitudes these singularities
(like all others) appear as poles in $\e$.  It may therefore
be most convenient to handle the initial-state singularities
using an electron structure function method~\cite{ElectronSF} 
implemented in the $\MSbar$ collinear factorization scheme.

In the next section we briefly describe our method for computing the
two-loop amplitudes.  Then we describe Catani's formula for the divergence
structure of the amplitudes, followed by a presentation of the finite
($\Ord(\e^0)$) terms for both $e^+e^-\to\mu^+\mu^-$ and Bhabha scattering. 
In the final section we give our conclusions,
including some discussion of the remaining ingredients still required 
for construction of a numerical program for Bhabha scattering at this
order.

\section{The Two-Loop Amplitudes}
\label{AmplitudesSection}

\begin{figure}[ht]
\centerline{\epsfxsize 4.2 truein \epsfbox{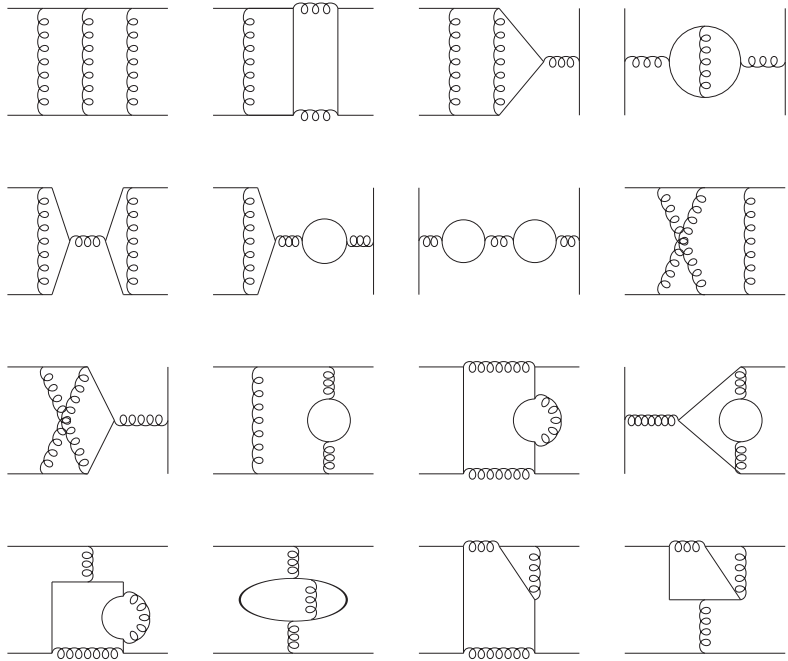}}
\vskip -.2 cm
\caption[]{
\label{DiagramsFigure}
\small The independent diagrammatic topologies for two-loop
four-fermion scattering in QED. }
\end{figure}

The 16 independent Feynman diagram topologies describing the two-loop
QED corrections to $e^+e^- \to \mu^+\mu^-$ and Bhabha scattering are
enumerated in \fig{DiagramsFigure}. In this figure we have suppressed
the fermion arrows.  After including the fermion arrows and distinct
labels for the external legs, there are a total of 47 Feynman
diagrams; however, many of these diagrams generate identical results.
Of the 47 diagrams, 35 contain no fermion loop, 11 contain one fermion
loop, and 1 contains two fermion loops.  The Bhabha amplitude may be
obtained from the $e^+e^- \to \mu^+\mu^-$ amplitude by adding to it
the same set of diagrams, but with an exchange of one pair of external
legs.  The $e^- \mu^- \to e^-\mu^-$ and $e^-\mu^+ \to e^-\mu^+$ amplitudes
may, of course, be obtained by crossing.

%

We have evaluated these diagrams interfered with the tree amplitudes and
summed over spins in the conventional dimensional regularization (CDR)
scheme.  This interference gives directly the two-loop virtual correction
to the $2 \to 2$ differential cross section.  
The rules for implementing CDR are straightforward
because all particle are treated uniformly in all parts of the
calculation.  In this scheme, all momenta and all Lorentz indices are
taken to be $D=4-2\eps$ dimensional vectors.  (The $\gamma$-matrices
remain as $4\times4$ matrices; i.e., ${\rm Tr}[1] = 4$.)

After performing all $\gamma$-matrix algebra present in the two-loop Feynman
diagrams, we use the conservation of momenta flowing on 
the internal lines to express the tensor structure of the diagrams in
terms of inverse scalar propagators and a small number of additional scalar
invariants containing loop momenta.  The inverse scalar propagators
cancel propagators in the denominator to generate simpler ``boundary''
integrals.  To handle the integrals containing scalar invariants,
we introduce Feynman parameters and interpret the resulting integrals
in terms of scalar integrals with multiple propagators, which are
then reduced to a set of master integrals with the help of equations in
refs.~\cite{PBReduction,NPBReduction,AGOTwo}.  

Proceeding in this way, we obtain an expression for the amplitude in 
terms of master integrals (of the type listed in ref.~\cite{NPBReduction}, 
plus a few more for the planar double box topology) multiplied by 
coefficient functions.  This expression is in principle valid to an arbitrary
order in $\e$, assuming that the master integrals could be evaluated 
to such an order.  However, it is a bit too lengthy to present here,
and for NNLO computations only the series expansion in $\e$ through 
$\Ord(\e^0)$ is required.
To carry out this expansion, we use expansions of the master integrals 
presented in 
refs.~\cite{PBScalar,NPBScalar,PBReduction,NPBReduction,AGOOne,AGOTwo}.
As noted in ref.~\cite{GloverTY}, there is a slight problem with the
original choice of basis~\cite{PBReduction} for the two master planar
double box integrals.  In that basis, the coefficients for generic 
tensor integrals contain $1/\e$ poles, necessitating an $\Ord(\e)$ 
evaluation of the master integrals.  Several solutions to this problem
have been presented~\cite{ReductionTalks,GloverTY}.  We have used a
slightly different solution, which is simply to use the original pair 
of master integrals defined in ref.~\cite{PBReduction}, except evaluated
in $d=6-2\e$ instead of $d=4-2\e$.  In $d=6-2\e$ the integrals have
neither ultraviolet nor infrared divergences, making them simpler
to evaluate through $\Ord(\e^0)$ than the $d=4-2\e$ integrals.

Many of the master integral expansions quoted in 
refs.~\cite{PBScalar,NPBScalar,PBReduction,NPBReduction,AGOOne,AGOTwo}
are in terms of Nielsen functions~\cite{NielsenRef}, 
$$
S_{n,p}(x) = {(-1)^{n+p-1} \over (n-1)! \, p!} 
\int_0^1 {dt \over t} \ln^{n-1} t \, \ln^p(1-xt)\,,
\equn\label{NielsenDef}
$$
with $n+p \leq 4$.  We have found it useful to express the results 
instead in terms of a minimal set of polylogarithms~\cite{Lewin},
$$
\eqalign{
\Li_n(x) &= \sum_{i=1}^\infty { x^i \over i^n }
          = \int_0^x {dt \over t} \Li_{n-1}(t), \cr
\Li_2(x) &= -\int_0^x {dt \over t} \ln(1-t) \,, \cr}
\equn\label{PolyLogDef}
$$
with $n=2,3,4$, using relations such as
$$
\eqalign{
S_{13}(x) &= - \Li_4(1-x) + \ln(1-x) \Li_3(1-x) 
+ {1\over2} \ln^2(1-x) \Bigl( \Li_2(x) - \zeta_2 \Bigr)
+ {1\over3} \ln^3(1-x) \ln x + \zeta_4 \,, \cr
S_{22}(x) &= \Li_4(x) - \Li_4(1-x) + \Li_4\biggl({-x\over1-x}\biggr)
- \ln(1-x) \Bigl( \Li_3(x) - \zeta_3 \Bigr)
\cr & \hskip1cm
+ {1\over24} \ln^4(1-x) - {1\over6} \ln^3(1-x) \ln x
+ {1\over2} \zeta_2 \ln^2(1-x) + \zeta_4 \,, 
\cr \hbox{for} & \quad 0 < x < 1. \cr}
\equn\label{SomeNielsenRelations}
$$
Here 
$$
\zeta_s \equiv \sum_{n=1}^\infty n^{-s} \,; \qquad \quad
\zeta_2 = {\pi^2\over6} \,, \qquad \zeta_3 = 1.202057\ldots, \qquad
\zeta_4 = {\pi^4\over90} \,.
\equn\label{ZetaValues}
$$

The analytic properties of the non-planar double box integrals are somewhat
intricate~\cite{NPBScalar}, since they are not real in any of the three 
kinematic channels for the $2 \to 2$ process,
$$
\eqalign{
&\hbox{$s$-channel}:   \quad \,  s > 0; \quad   t, \, u < 0\,, \cr
&\hbox{$t$-channel}:   \quad \;  t \,> 0; \quad   s, \, u < 0\,, \cr
&\hbox{$u$-channel}\!: \quad \,  u > 0; \quad    s, \; t < 0\,, \cr}
\equn\label{ChannelDefs}
$$
where $s=(k_1+k_2)^2$, $t=(k_1-k_4)^2$, and $u=(k_1-k_3)^2$.
Therefore we shall present explicit formulae for the finite terms in the
amplitude in both the $s$- and $u$-channels; those in the $t$-channel
will be related by symmetries.


\subsection{General Structure of Divergences}
\label{DivergenceSubsection}

Dimensionally regulated two-loop amplitudes for four massless fermions 
contain poles in $\e=(4-d)/2$ up to $1/\e^4$.  The structure of most of 
these singularities has already been exposed by Catani~\cite{Catani}, who
described the infrared behavior of general two-loop QCD processes.  
We shall therefore adopt his notation in presenting our results.

We work with ultraviolet renormalized amplitudes, and employ the
$\MSbar$ running coupling for QED, $\alpha(\mu^2)$.  Of course this
scheme can always be converted to another one, for example
$\alpha(Q^2)$ defined via the photon propagator at momentum transfer
$Q$, by a finite renormalization.  The relation between the bare
coupling $\alpha^u$ and $\alpha(\mu^2)$ through two-loop order can be
expressed as~\cite{Catani}
$$
\alpha^u \, \mu_0^{2\e} \, S_\e \ =\ 
\alpha(\mu^2) \, \mu^{2\e} \Biggl[ 1 - \alpha(\mu^2) { \beta_0 \over \e}
   + \alpha^2(\mu^2) \biggl( { \beta_0^2 \over \e^2} 
                           - {\beta_1 \over 2\e} \biggr) 
   + \Ord(\alpha^3(\mu^2)) \Biggr] \, ,
\equn\label{TwoloopCoupling}
$$
where $S_\e = \exp[\e (\ln4\pi + \psi(1))]$ and 
$\gamma = -\psi(1) = 0.5772\ldots$ is Euler's constant.
The first two coefficients of the QED beta function are
$$
  \beta_0 = - {N_{\! f} \over 3\pi}\,, \qquad
  \beta_1 = - {N_{\! f} \over 4\pi^2}\,,
\equn\label{QEDBetaCoeffs}
$$
where $N_{\! f}$ is the number of light (massless) charge 1 fermions.

The renormalized four-fermion amplitude is expanded as
$$
\eqalign{
\cm_4(\alpha(\mu^2),\mu^2;\{p\}) &=  4\pi\alpha(\mu^2) \, 
\Biggl[ \cm_4^{(0)}(\mu^2;\{p\}) 
+ { \alpha(\mu^2) \over 2\pi } \cm_4^{(1)}(\mu^2;\{p\}) \cr
& \hskip3cm
+ \biggl( { \alpha(\mu^2) \over 2\pi } \biggr)^2 \cm_4^{(2)}(\mu^2;\{p\})
+ \Ord(\alpha^3(\mu^2))
\Biggr] \,. \cr}
\equn\label{RenExpand}
$$

The infrared divergences of a renormalized two-loop amplitude in QCD or
QED are~\cite{Catani},
$$
\eqalign{
| \cm_n^{(2)}(\mu^2; \{p\}) \ra_{\RS} &= 
{\bom I}^{(1)}(\e, \mu^2; \{p\}) 
\; | \cm_n^{(1)}(\mu^2; \{p\}) \ra_{\RS} \cr
& \hskip 0.3 cm 
+ {\bom I}^{(2)}_{\RS}(\e, \mu^2; \{p\}) \; 
  | \cm_n^{(0)}(\mu^2; \{p\}) \ra_{\RS}
+ | \cm_n^{(2){\rm fin}}(\mu^2; \{p\}) \ra_{\RS} \,, \cr}
\equn\label{TwoloopCatani}
$$
where $|\cm_n^{(L)}(\mu^2; \{p\}) \ra_{\RS}$ is a color space
vector representing the renormalized $L$ loop amplitude.   
The subscript $\RS$ stands for the choice of renormalization scheme,
and $\mu$ is the renormalization scale. These color space vectors give the
amplitudes via,
$$
{\cal M}_n(1^{a_1},\dots,n^{a_n}) \equiv
\la a_1,\dots,a_n \,
| \, \cm_n(p_1,\ldots,p_n)\ra \,,
\equn\label{MnVec}
$$
where the $a_i$ are color indices.  The divergences of ${\cal M}_n$
are encoded in the color operators 
${\bom I}^{(1)}(\e,\mu^2;\{p\})$ and 
${\bom I}^{(2)}(\e,\mu^2;\{p\})$.  In the QED case, 
the color space language is clearly unnecessary; $\cm_n^{(L)}$
and ${\bom I}^{(L)}$ are just numbers.

In QCD, the operator ${\bom I}^{(1)}(\e,\mu^2;\{p\})$ is given by
$$
{\bom I}^{(1)}(\e,\mu^2;\{p\}) =  \frac{1}{2}
{e^{-\e \psi(1)} \over \Gamma(1-\e)} \sum_{i=1}^n
\, \sum_{j \neq i}^n \, 
{\bom T}_i \cdot {\bom T}_j \biggl[ {1 \over \e^2}
 + {\gamma_i \over {\bom T}_i^2 } \, {1 \over \e} \biggr] 
\biggl( \frac{\mu^2 e^{-i\lambda_{ij} \pi}}{2 p_i\cdot p_j} \biggr)^{\e}
 \,,
\equn\label{CataniGeneral}
$$
where $\lambda_{ij}=+1$ if $i$ and $j$ are both incoming or outgoing
partons and $\lambda_{ij}=0$ otherwise. The color charge ${\bom T}_i =
\{T^a_i\}$ is a vector with respect to the generator label $a$, and an
$SU(N_c)$ matrix with respect to the color indices of the outgoing 
parton $i$.  The values required for QCD are
$$
\eqalign{
{\bom T}_q^2 & = {\bom T}_\qb^2 = C_F \,,  \hskip 2 cm 
{\bom T}_g^2 = C_A = N_c \,,  \cr
\gamma_q & = \gamma_{\qb} = {3 \over 2} C_F \,,  \hskip 2 cm 
\gamma_g = {11\over 6}\, C_A - {2 \over 3} T_R \, N_{\! f} \,. \cr} 
\equn\label{QCDValues}
$$

For QED we let $C_A \to 0$, $C_F \to 1$, $T_R \to 1$ and 
${\bom T}_i \cdot {\bom T}_j \to e_i e_j = \pm 1$, where the $e_i$ 
are the electric charges, to obtain
$$
{\bom I}^{(1)}(\e,\mu^2;\{p\}) =
{e^{-\e \psi(1)} \over \Gamma(1-\e)} 
\biggl( {2 \over \e^2} + {3 \over \e} \biggr)
\Biggl[
-\biggl({\mu^2\over -s} \biggr)^{\e} 
-\biggl({\mu^2\over -t} \biggr)^{\e} 
+\biggl({\mu^2\over -u} \biggr)^{\e} \Biggr] \,,
\equn\label{Catani1QED}
$$
for the four-fermion amplitude 
$$
  e^+(k_1) \, e^-(k_2)\ \to\ \mu^+(k_4) \, \mu^-(k_3) \,.
\equn\label{eemumuKin}
$$
(Note that the charges of incoming states should be reversed in computing
${\bom T}_i \cdot {\bom T}_j$.)

The operator ${\bom I}^{(2)}_{\RS}$ is given in QCD by~\cite{Catani}
$$
\eqalign{
{\bom I}^{(2)}_{\RS}(\e,\mu^2;\{p\}) 
& = - \frac{1}{2} {\bom I}^{(1)}(\e,\mu^2;\{p\})
\left( {\bom I}^{(1)}(\e,\mu^2;\{p\}) + {4 \pi \beta_0 \over \e} \right)
 \cr
& \hskip1cm
+ {e^{+\e \psi(1)} \Gamma(1-2\e) \over \Gamma(1-\e)} 
\left( {2\pi\beta_0 \over \e} + K \right) {\bom I}^{(1)}(2\e,\mu^2;\{p\})
\cr
& \hskip1cm
 + {\bom H}^{(2)}_{\RS}(\e,\mu^2;\{p\}) \, , \cr}
\equn\label{CataniGeneralI2}
$$
where the coefficient $K$ is:
$$
K = \left( \frac{67}{18} - \frac{\pi^2}{6} \right) C_A 
 - \frac{10}{9} T_R N_{\! f} \;\;.
\equn\label{CataniK}
$$
For the QED process~(\ref{eemumuKin}), we insert ${\bom I}^{(1)}$ 
from~\eqn{Catani1QED}, take $\beta_0$ and $\beta_1$ 
from \eqn{QEDBetaCoeffs}, and let $K \to - 10 N_{\! f}/9$.

The function ${\bom H}^{(2)}_{\RS}$ is process-dependent but has only
{\em single} poles:
$$
 {\bom H}^{(2)}_{\RS}(\e,\mu^2;\{p\}) \ =\ \Ord(1/\e) \,.
\equn\label{CataniGenH}
$$
Ref.~\cite{Catani} does not give an expression for ${\bom H}^{(2)}_{\RS}$
for a general amplitude, but only for the case of a $q\bar{q}$ pair,
i.e. a single charged fermion pair.  The result, which is extracted from 
the two-loop QCD computation of the electromagnetic form factor of the 
quark~\cite{QuarkFormFactor}, is
$$
\eqalign{
 {\bom H}^{(2)}_{q\bar{q},\CDR}(\e,\mu^2;\{p\})
&= {1 \over 4\e} {e^{-\e \psi(1)} \over \Gamma(1-\e)}
 \Biggl( { \mu^2 e^{-i\lambda_{12}\pi} \over 2p_1\cdot p_2 } \Biggr)^{2\e}
 \biggl[ {1\over4} \gamma_{(1)} 
       + 3 C_F K 
       + 5 \zeta_2 \pi \beta_0 C_F 
       - {56\over9} \pi \beta_0 C_F \cr
& \hskip5cm
       - \biggl( {16\over9} - 7 \zeta_3 \biggr) C_F C_A \biggr] \ , \cr}
\equn\label{CataniHqq}
$$
where
$$
\gamma_{(1)}\ =\ 
        ( - 3 + 24 \zeta_2 - 48 \zeta_3 ) C_F^2
+ \biggl( -{17\over3} - {88\over3} \zeta_2 + 24 \zeta_3 \biggr) C_F C_A
+ \biggl( {4\over3} + {32\over3} \zeta_2 \biggr) C_F T_R N_{\! f} \,.
\equn\label{Catanigamma1}
$$
Performing the usual conversion to QED yields a result applicable to 
the electromagnetic form factor of the electron,
$$
 {\bom H}^{(2)}_{e^+e^-,\CDR}(\e,\mu^2;\{p\})
= {1\over4\e} {e^{-\e \psi(1)}\over\Gamma(1-\e)}
 \Biggl( { \mu^2 e^{-i\lambda_{12}\pi} \over 2p_1\cdot p_2 } \Biggr)^{2\e}
 \biggl[ -{3\over4} + 6 \zeta_2 - 12 \zeta_3
         + \biggl( - {25\over27} + \zeta_2 \biggr) N_{\! f} \biggr] \, .
\equn\label{CataniHee}
$$

Using our two-loop computation, and an all-orders-in-$\e$ computation
of the one-loop amplitude for $e^+e^-\to\mu^+\mu^-$ 
(see sect.~\ref{eemumuOneloopSubsection}),
we have verified that the singular behavior of the 
$e^+e^- \to \mu^+\mu^-$ amplitude in CDR agrees precisely with
that predicted by~\eqn{TwoloopCatani} in all three kinematic 
channels.\footnote{Strictly speaking, we have
computed the interference of the two-loop amplitude with the
tree-amplitude, summed over intermediate fermion spins, so in our
verification \eqn{TwoloopCatani} should be similarly understood to be
interfered with the tree amplitude.}\ 
In addition, we have extracted the function 
${\bom H}^{(2)}_{e^+e^-\mu^+\mu^-,\CDR}$
controlling the $1/\e$ poles in \eqn{TwoloopCatani}.  We obtain
$$
\eqalign{
 {\bom H}^{(2)}_{e^+e^-\mu^+\mu^-,\CDR}(\e,\mu^2;\{p\})
&= {1\over4\e} {e^{-\e \psi(1)}\over\Gamma(1-\e)}
 \times 2 \, \Biggl[
 \biggl({\mu^2\over -s} \biggr)^{2\e} 
+\biggl({\mu^2\over -t} \biggr)^{2\e} 
-\biggl({\mu^2\over -u} \biggr)^{2\e} \Biggr] \cr
& \hskip2.7cm
 \times \biggl[ -{3\over4} + 6 \zeta_2 - 12 \zeta_3
         + \biggl( - {25\over27} + \zeta_2 \biggr) N_{\! f} \biggr] \, . \cr}
\equn\label{UsHeemumu}
$$
This result agrees with a ``naive'' generalization from the form factor case,
in which one sums \eqn{CataniHee} over the six pairs of charged legs
in the four-point amplitude, weighted by the sign of the charge product 
$e_ie_j$.  (Note that the factors of $(\mu^2/(-s_{ij}))^{2\e}$ are purely 
conventional here, since their deviation from unity
only contributes at the level 
of finite parts, $\Ord(\e^0)$.  However, the overall normalization is
predicted correctly by the sum over the six pairs.)

\subsection{$e^+e^- \to \mu^+\mu^-$ at One Loop to All Orders in $\e$}
\label{eemumuOneloopSubsection}

In order to verify the structure of the infrared singularities, and to
extract the finite remainder of the two-loop amplitude presented below, we
computed the one-loop $e^+e^- \to \mu^+\mu^-$ amplitude (interfered
with the tree amplitude) to all orders in $\e$.  The result is
$$
 \sum_{\rm spins} \cm_4^{(1)} \cm_4^{(0)\,\dagger} 
 =  {2\over3} {N_{\! f} \over \e} 
        \sum_{\rm spins} \cm_4^{(0)} \cm_4^{(0)\,\dagger}
    + \biggl[ A^{(1)} + S\Bigl[ A^{(1)} \Bigr] \biggr] 
  \bigg\vert_{\xi=1} \, ,
\equn\label{OneLoopAllOrders1}
$$
where the first term is the $\MSbar$ counterterm, expressed in terms
of the tree-level interference
$$
 \sum_{\rm spins} \cm_4^{(0)} \cm_4^{(0)\,\dagger} 
 =  8 \biggl[ { t^2+u^2 \over s^2 } \, - \, \e \biggr] \ , 
\equn\label{CDRTree}
$$
and
$$
\eqalign{
A^{(1)} &= 
  4 \xi  (1-2\e) {u \over s^2} 
    \Bigl[ (2-3\e) u^2 - 6 \e t u + 3(2-\e) t^2 \Bigr] 
          \Boxsixint(s,t) \cr
& \hskip1cm
- 4 {\xi \over 1-2\e} {t \over s^2} 
    \Bigl[ (4-12\e+7\e^2) t^2 - 6\e (1-2\e) t u + (4-10\e+5\e^2) u^2 \Bigr]
           \Triint(t) \cr
& \hskip1cm
- {8 \over (1-2\e)(3-2\e)} {1\over s} \biggl[ 
  2\e (1-\e) \, t ( (1-\e) t - \e u ) \, N_{\! f}
  - \e (3-2\e) (2-\e+2\e^2) t u    \cr
& \hskip5cm
  + (1-\e) (3-2\e) ( 2 - (1-\xi) \e + 2 \e^2 ) t^2 \biggr]
           \Triint(s)\,.
  \cr}
\equn\label{OneLoopAllOrders2}
$$
The symmetry operation $S$ acts as
$$
 S: \quad t \lr u, \qquad \xi \lr -\xi.
\equn\label{Sdef}
$$
After carrying out the operation of $S$, one should then set $\xi=1$.
(Basically, $\xi$ allows us to separate diagrams based on whether they 
have an even or odd number of photons attached to the muon line.  Because 
photons have ${\rm C} = -1$, this criterion governs the $t \lr u$ 
symmetry properties.)

In \eqn{OneLoopAllOrders2}, $\Boxsixint(s,t)$ and $\Triint(s)$ are
one-loop box and triangle integrals, the former evaluated in an
expansion around $d=6-2\e$.  For the divergence
formula~(\ref{TwoloopCatani}), we need their series expansions in $\e$ 
through $\Ord(\e^2)$.  In the $u$-channel, where the functions are 
manifestly real, their expansions are given by
$$
\eqalign{
\Boxsixint(s,t) &= 
{ u^{-1-\e} \over 2 (1-2\e) } 
     \biggl( 1 - {\pi^2\over12} \e^2 \biggr)  \Biggl[ 
   {1\over2} \Bigl( (V-W)^2 + \pi^2 \Bigr)
 + 2 \e  \biggl( \li3(-v) - V \li2(-v) - {1\over3} V^3 
                                         - {\pi^2\over2} V \biggr)
\cr & \hskip2.5cm
 - 2 \e^2 \biggl( \li4(-v) + W \li3(-v) - {1\over2} V^2 \li2(-v)
             - {1\over8} V^4 - {1\over6} V^3 W + {1\over4} V^2 W^2
\cr & \hskip3.5cm
             - {\pi^2\over4} V^2 - {\pi^2\over3} V W - 2 \zeta_4 \biggr)
 \ +\ \hbox{$(s \lr t)$}\ \Biggr]\ +\ \Ord(\e^3), \cr
 \Triint(s) &= - { (-s)^{-1-\e} \over \e^2 } \biggl[ 
   1 - {\pi^2\over12} \, \e^2 - {7\over3} \zeta_3 \, \e^3
     - {47\over16} \zeta_4 \, \e^4 \biggr]\ + \Ord(\e^3)\,,
 \cr}
\equn\label{OneLoopIntegralExp}
$$
where
$$
 v = {s\over u} \, , \quad w = {t\over u} \, , \quad 
 V = \ln\biggl(-{s\over u}\biggr) \, , \quad 
 W = \ln\biggl(-{t\over u}\biggr) \, .
\equn\label{uChannelvwVWdef}
$$
The expansions in the $s$- and $t$-channels can be found using
analytic continuation formulae such as~\cite{AGOTwo}
$$
\eqalign{
  \ln(1-x+i\varepsilon) &= \ln(x-1) + i\pi \,, \cr
  \li2(x+i\varepsilon) &= - \li2\biggl({1\over x}\biggr)
    - {1\over2} \ln^2 x + {\pi^2\over3} + i \pi \ln x \,, \cr
  \li3(x+i\varepsilon) &= \li3\biggl({1\over x}\biggr)
    - {1\over6} \ln^3 x + {\pi^2\over3} \ln x + i {\pi\over2} \ln^2 x \,, \cr
  \li4(x+i\varepsilon) &= - \li4\biggl({1\over x}\biggr)
    - {1\over24} \ln^4 x + {\pi^2\over6} \ln^2 x + 2 \zeta_4
    + i {\pi\over6} \ln^3 x \,, \cr
 x &> 1 \,, \cr}
\equn\label{polylogcontinue}
$$
where $i\varepsilon$ is an imaginary infinitesimal added 
to $s,t$ or $u$ before continuing.

We have verified that through $O(\eps^0)$ our result for the 
one-loop amplitude agrees with a previous 
calculation~\cite{BeenakkerBhabha}, up to terms which can be identified
as being due to the conversion between dimensional regularization 
and a photon mass regularization.

\subsection{Modifications for Bhabha Scattering}

In comparison with the process $e^+e^- \to \mu^+\mu^-$ described above,
the Bhabha scattering process
$$
  e^+(k_1) \, e^-(k_2)\ \to\ e^+(k_4) \, e^-(k_3) \,,
\equn\label{BhabhaKin}
$$
has additional exchange diagrams.  In general, the interference required
for Bhabha scattering is given by
$$
\eqalign{
 \sum_{\rm spins} \cm_4^{(L_1)} \cm_4^{(L_2)\,\dagger} 
  \bigg\vert_{\rm Bhabha}
\ &=\ 
   \sum_{\rm spins} \cm_4^{(L_1)} \cm_4^{(L_2)\,\dagger} 
+  \sum_{\rm spins} \cm_4^{(L_1)} \cmt_4^{(L_2)\,\dagger} 
\cr & \hskip0.1cm
+ U \Biggl[  \sum_{\rm spins} \cm_4^{(L_1)} \cm_4^{(L_2)\,\dagger} 
+ \sum_{\rm spins} \cm_4^{(L_1)} \cmt_4^{(L_2)\,\dagger} \Biggr] \ , \cr}
\equn\label{Bhabhapermsum}
$$
where the symmetry $U$ acts as
$$
 U: \quad s \lr t,
\equn\label{Udef}
$$
$\cm_4^{(L)}$ is the $L$-loop amplitude for $e^+e^- \to \mu^+\mu^-$,
and $\cmt_4^{(L)}$ is the same $L$-loop amplitude but with legs 1 and 3
interchanged (taking into account the Fermi statistics minus sign).

\subsection{Bhabha Scattering at One Loop to All Orders in $\e$}
\label{BhabhaOneloopSubsection}

In the CDR scheme, the tree-level exchange contribution required for 
Bhabha scattering in~\eqn{Bhabhapermsum} is
$$
 \sum_{\rm spins} \cm_4^{(0)} \cmt_4^{(0)\,\dagger} 
 =  8 \, (1-\e) \biggl[ {u^2 \over st} + \e \biggr] \, .
\equn\label{CDRcrossTree}
$$
The one-loop exchange contribution, evaluated to all orders in $\e$, is
given by
$$
 \sum_{\rm spins} \cm_4^{(1)} \cmt_4^{(0)\,\dagger} 
 =  {2\over3} {N_{\! f} \over \e} 
        \sum_{\rm spins} \cm_4^{(0)} \cmt_4^{(0)\,\dagger}
    + \tilde{A}^{(1)} \ ,
\equn\label{OneLoopExchangeAllOrders1}
$$
where
$$
\eqalign{
\tilde{A}^{(1)} &= \ 
  8 (1-2 \e) {u \over s t}  
      \Bigl( (1-4\e+\e^2) t^2 - 2 \e (2-\e) t u + (1-\e)^2 u^2 \Bigr)
      \Boxsixint(s,t) 
\cr & \hskip0.3cm
+ 8 (1-2 \e) {1\over s}   
    \Bigl( \e (2-3 \e-\e^2) t^2 + 2 \e (1-3\e-\e^2) t u
         - (2-2\e+3\e^2+\e^3) u^2 \Bigr)
      \Boxsixint(s,u)
\cr & \hskip0.3cm
- { 8 (1-\e) \over (1-2 \e) (3-2 \e) } {1\over t}   
     \biggl[  2 \e (1-\e) (u^2 + \e s t) N_{\! f}
\cr & \hskip3.8cm
         - (3-2 \e) \Bigl( 2 \e (1+\e^2) t^2 + \e (3+2 \e^2) t u 
                         - 2 (1-\e+\e^2) u^2 \Bigr) \biggr]
      \Triint(s)
\cr & \hskip0.3cm
+ {8 \over 1-2 \e} {1\over s}  \Bigl( \e (2-5\e+2\e^2-\e^3) t^2 
                                    + \e (1-3\e+\e^2-\e^3) t u
                                    - (1-\e) (2-3\e-\e^2) u^2 \Bigr)
      \Triint(t)
\cr & \hskip0.3cm
- {8 \over 1-2 \e} {u\over s t}  \Bigl( \e (2-4\e+\e^2-\e^3) t^2 
                                      + \e (2-3 \e-\e^3) t u
                                      - (1-\e) (2-4\e-\e^2) u^2 \Bigr)
      \Triint(u)  \, . 
\cr}
\equn\label{OneLoopExchangeAllOrders2}
$$

Using these results, and the computation of the two-loop exchange terms, 
we again find that the additional singular terms in Bhabha scattering 
are described by \eqn{TwoloopCatani}, where (not surprisingly) 
${\bom H}^{(2)}_{\RS}$ is given by precisely the same 
expression~(\ref{UsHeemumu}) that we found for $e^+e^- \to \mu^+\mu^-$.

\subsection{Finite Contributions to the Amplitudes}

\subsubsection{$e^+e^-\to\mu^+\mu^-$}

Finally we give the real (dispersive) part of the finite remainder 
in \eqn{TwoloopCatani}, interfered with the tree amplitude in the CDR
scheme.  First we treat the $e^+e^-\to\mu^+\mu^-$ process~(\ref{eemumuKin}).
It is convenient to decompose the finite part according to
the number of light flavors, $N_{\! f}$,
$$
  \sum_{\rm spins}
\Re \Bigl[ \cm_4^{(2){\rm fin}} \cm_4^{(0)\,\dagger} \Bigr]\ =\ 
 8 \Bigl[ F^{(0)} + N_{\! f} \, F^{(1)} + N_{\! f}^2 \, F^{(2)} \Bigr]\,. 
\equn\label{TwoloopRemainder}
$$
In the $s$-channel, the functions $F^{(i)}$ are given by 
$$
F^{(i)}\ =\ \biggl[ F_s^{(i)} + S \Bigl[ F_s^{(i)} \Bigr] \biggr] 
  \bigg\vert_{\xi=1} \ ,
\equn\label{FfromFs}
$$
where
$$
\eqalign{
 F_s^{(0)}\
 &=\ 2 {x^3\over y} X^2  
+ (x^2+y^2) \Biggl[ 4 (2-\xi) \Bigl( \li4(-x) - X \li3(-x) 
                              + {1\over2} X^2 \li2(-x) \Bigr)
      + {4\over3} \xi \pi^2 \li2(-x)
\cr & \hskip2.0cm
  + \biggl( {1\over6} (1+\xi) X^3 + {2\over3} (2-\xi) X^2 Y 
          - {1\over2} X Y^2 - 3 (X-Y) X
          - 2 {\pi^2\over3} \Bigl( (3+\xi) X - 3 Y \Bigr)
\cr & \hskip2.3cm
          + {1\over2} (11-16 \xi) X - {9\over2} Y 
          + \xi \Bigl( - 12 \zeta_3 + \pi^2 + {93\over4} \Bigr) \biggr) X 
    - {43\over2} \zeta_4 - {15\over 2} \zeta_3 
    + {29\over24} \pi^2 + {511\over32} \Biggr]
\cr & \hskip0.5cm
+ (x-y) \Biggl[ 8 \li4(-x/y)
          + 6 (2+\xi) \li4(-x) 
          + ( 4 (1-\xi) X - 12 Y ) \li3(-x)
\cr & \hskip2.0cm
          - \biggl( (2-\xi) X^2 - 4 X Y + {4\over3} \pi^2 \biggr) \li2(-x)
          - (2+\xi)  \Bigl( \li3(-x) - X \li2(-x) \Bigr)
\cr & \hskip2.0cm
          + \biggl( - {1\over12} (6+\xi) X^3 + {2\over3} X^2 Y 
                    + {1\over6} (1+4 \xi) \pi^2 X
                    + {1\over6} (10+\xi) X^2 - {1\over2} (2-\xi) X Y 
\cr & \hskip2.5cm
                    + {1\over2} (1+6\xi) X
                    - 4 (2-\xi) \zeta_3 - (1+4\xi) {\pi^2\over6} 
                    - 6 \xi \biggr) X
      + \xi \biggl( - 6 \zeta_4 - 2 \zeta_3 + 2 {\pi^2\over3} \biggr)
   \Biggr]
\cr & \hskip0.5cm
   + (2-\xi) \Bigl( \li3(-x) - X \li2(-x) \Bigr)
   + \biggl( {1\over6} (5-3 \xi) X^2 - {1\over2} (3-\xi) X Y 
            - (1-4\xi) {\pi^2\over6} \biggr) X
\cr & \hskip0.5cm
   + \biggl( - {1\over2} (1+6 \xi) X - {1\over2} Y - 6\xi \biggr) X 
   - 4 \zeta_3 + {\pi^2\over3} \, , \cr}
\equn\label{Fs0}
$$
$$
\eqalign{
F_s^{(1)}\ 
&=\ {1\over9} \Biggl\{ 
   (x^2+y^2) \biggl( \xi \Bigl[
           12 \Bigl( \li3(-x) - X \li2(-x) \Bigr)
         + \Bigl( 4 X^2 - 6 X ( Y + 2 \ln(\mu^2/s) ) + 20 \pi^2
\cr & \hskip3.0cm
         - 29 X + 36 \ln(\mu^2/s) + 33 \Bigr) X \Bigr]
         + {87\over2} \ln(\mu^2/s) + {35\over2} \zeta_3 
         + {7\over4} \pi^2 + {685\over9} \biggr) 
\cr & \hskip1cm
  - 2 \xi (x-y) \biggl( \Bigl( X^2 - 3 (X - 1) \ln(\mu^2/s)
               - {13\over2} X + 4 \pi^2 + 8 \Bigr) X  -  2 \pi^2 \biggr)
\cr & \hskip1cm
  - \xi ( 3 X + 6 \ln(\mu^2/s) + 16 ) X \Biggr\} \, ,\cr}
\equn\label{Fs1}
$$
$$
\eqalign{
\hskip-7cm
F_s^{(2)}\ &=\ {4\over9} x^2  
    \biggl[ \ln^2(\mu^2/s) + {10\over3} \ln(\mu^2/s) - \pi^2 + {25\over9}
    \biggr]\, , 
\cr}
\equn\label{Fs2}
$$
with
$$
 x = {t\over s} \, , \quad y = {u\over s} \, , \quad 
 X = \ln\biggl(-{t\over s}\biggr) \, , \quad 
 Y = \ln\biggl(-{u\over s}\biggr) \, ,
\equn\label{sChannelxyXYdef}
$$
and the symmetry operation $S$ is given in \eqn{Sdef}.

In the $u$-channel, the $F^{(i)}$ are given by 
$$
F^{(i)}\ =\ F_u^{(i)} \Big\vert_{\xi=1} \ ,
\equn\label{FfromFu}
$$
where 
$$
\eqalign{
 F_u^{(0)}\ &=
\ 2 {x-y \over y} \Bigl( (V-W)^2 + \pi^2 \Bigr)
              - 2 (x-y) \biggl( {1\over x} - 3 \xi \biggr) V^2 
\cr & \hskip0.5cm
+ (x^2+y^2) \Biggl[
  - 4 \biggl[ (2+\xi) \Bigl( \li4(-v) - V \li3(-v) 
                             + {1\over2} V^2 \li2(-v) \Bigr)
\cr & \hskip3.2cm
             + (2-\xi) \biggl( \li4(-v/w) 
             + (V-W) \Bigl( \li3(-v) + \li3(-w) - W \li2(-w) 
\cr & \hskip3.2cm
                     - {1\over2} (V+W) \li2(-v) \Bigr) \biggr)
     \biggr]
  + 2 (6+\xi) {\pi^2\over3} \li2(-v) 
\cr & \hskip2.5cm
  - {4\over3} \xi V^3 W + (4-\xi) V^2 W^2 
  - {2\over3} (7-2\xi) V W^3 - {1\over6} (1-2\xi) W^4 
\cr & \hskip2.5cm
  + 6 V W^2 - 3 W^3
  + 16 \xi V W + {1\over2} (9-16 \xi) W^2
  + {93\over4} \xi W 
\cr & \hskip2.5cm
  + \pi^2 \biggl( - 2 V^2 + {2\over3} (3+\xi) V W 
                  - {1\over3} (3-\xi) W^2 
                  - 6 V - (3-\xi) W  \biggr) 
\cr & \hskip2.5cm
  + 4 \Bigl( (2-\xi) V - 2 (1+\xi) W \Bigr) \zeta_3
  + 34 \zeta_4 - 15 \zeta_3 - (25+96 \xi) {\pi^2\over12} + {511\over16} 
  \Biggr]
\cr & \hskip0.5cm
- (x-y) \Biggl[ 
    6 \Bigl( (2+\xi) \li4(-v/w) - (2-\xi)   \li4(-v) \Bigr)
  - 16 \li4(-w) 
\cr & \hskip2.2cm
  + 4 \Bigl( (1-\xi)   W + (2+\xi)   V \Bigr) \li3(-w)
  + 4 \Bigl( 4 V - (2+\xi) W \Bigr)   \li3(-v)
\cr & \hskip2.2cm
  - (2+\xi) \Bigl( \li3(-w) + W   \li2(-v) \Bigr)
  - 4 \Bigl( \li3(-v) - V   \li2(-v) \Bigr)
\cr & \hskip2.2cm
  + \biggl( - 4 V^2 + 2 (2+\xi) V W + (2-\xi) W^2 
      - (10+3 \xi) {\pi^2\over3} \biggr) \li2(-v) 
\cr & \hskip2.2cm
  + {1\over3} \xi V^4 - {2\over3} (2+\xi) V^3 W + {5\over2} (2+\xi) V^2 W^2 
  - {1\over3} (4+7\xi) V W^3 + {1\over3} (2+\xi) W^4
\cr & \hskip2.2cm
  + {2\over3} \xi V^3 - (2+\xi) V^2 W + 2 V W^2 - {1\over6} (10+\xi) W^3
  + (5+6 \xi) V W - {1\over2} (5+6 \xi) W^2
\cr & \hskip2.2cm
  + {\pi^2\over6} \Bigl( 2 (6-\xi) V^2 - 2 (13+4\xi) V W + (5+3\xi) W^2
               + 2 (12-\xi) V - (3-2\xi) W \Bigr)
\cr & \hskip2.2cm
  - 4 \Bigl( (2-\xi) V - W \Bigr) \zeta_3
  - 6 \xi (2 V - W) 
  + (121+3\xi) \zeta_4 + (2+5\xi) \zeta_3 - (15+26\xi) {\pi^2\over6} 
  \Biggr]
\cr & \hskip0.5cm  
  + 2 \xi \Bigl( \li3(-v) - V \li2(-v) - V^2 W \Bigr)
  - (2-\xi) \Bigl( \li3(-w) - W   \li2(-w) \Bigr)
\cr & \hskip0.5cm  
  - (1-\xi) V W^2 + {1\over6} (5-3 \xi) W^3
  + 6 \xi V W + {1\over2} (1-6 \xi) W^2 - 6 \xi W
\cr & \hskip0.5cm  
  + {\pi^2\over6} \Bigl( 2 (4+3 \xi) V + (8-5 \xi) W \Bigr)
  - (6+\xi) \zeta_3 + (1-18 \xi) {\pi^2\over6} \,,
 \cr}
\equn\label{Fu0}
$$
$$
\eqalign{
F_u^{(1)} \
 &=\ {1\over9} \Biggl\{ 
  (x^2+y^2) \Biggl[  - \xi \biggl[ 
           12 \biggl( \li3(-w) - W   \li2(-w)
             + 2   ( \li3(-v) - V   \li2(-v) )
\cr & \hskip3.8cm  
             - {3\over2} V^2 W + {1\over2} V W^2 - {1\over3} W^3 
             + ( W^2 - 2 V W - 3 W + \pi^2 ) \ln(\mu^2/(-s)) \biggr)
\cr & \hskip3.8cm  
        + 29 (W - 2 V) W - 33 W + 2 \pi^2 (3 V - 4 W) \biggr]
\cr & \hskip2.7cm  
      + 87 \ln(\mu^2/(-s)) 
      + (35+12\xi) \zeta_3 + (7-58\xi) {\pi^2\over2} + {1370\over9} \Biggr]
\cr & \hskip0.5cm  
  + \xi (x-y) \Biggl[ 
        2 V^3 + 2 (V-W)^3
      + 13 ( 2 (V-W) V + W^2 )
\cr & \hskip2.5cm  
      + 6 \Bigl( 2 (V-W) V + W^2 + 2 V - W + \pi^2 \Bigr)   \ln(\mu^2/(-s)) 
      + 16 (2 V-W) 
\cr & \hskip2.5cm  
- 2 \pi^2 (V+W) + 9 \pi^2 \Biggr]
\cr & \hskip0.5cm  
  - \xi \biggl[ \Bigl( 3  W - 6  V + 6 \ln(\mu^2/(-s)) + 16 \Bigr) W 
              + 3 \pi^2 \biggr] \Biggr\} \, ,
     \cr}
\equn\label{Fu1}
$$
$$
\hskip-6cm
F_u^{(2)} \ =\ 
{4\over9} (x^2+y^2) 
    \biggl[ \ln^2(\mu^2/(-s)) + {10\over3} \ln(\mu^2/(-s)) + {25\over9}
    \biggr] \, .
\equn\label{Fu2}
$$
Here $x$, $y$ are defined in \eqn{sChannelxyXYdef}, 
whereas $v$, $w$, $V$, $W$ are defined in \eqn{uChannelvwVWdef}.

In the $t$-channel, the functions $F^{(i)}$ are given by the 
action of the symmetry $S$ of \eqn{Sdef} on the $u$-channel results,
$$
F^{(i)}\ =\ S \Bigl[ F_u^{(i)} \Bigr] \Big\vert_{\xi=1} \, .
\equn\label{FtfromFu}
$$

The two-loop virtual contribution to the $e^+e^- \to \mu^+\mu^-$
unpolarized cross section, restoring overall factors and averaging over
initial spins, is given by
$$
{d \sigma^{(2)} \over dt}
\ =\ {1 \over 16 \pi s^2 } \times
  {(4\pi\alpha)^2 \over 4} \times
 \Bigl({\alpha\over2\pi}\Bigr)^2 \times
 2 \, \sum_{\rm spins} \Re \Bigl[ \cm_4^{(2)} \cm_4^{(0)\,\dagger} \Bigr] \,.
\equn\label{FullCrossSection}
$$

\subsubsection{Bhabha Scattering}

For the finite two-loop remainder for the Bhabha scattering 
process~(\ref{BhabhaKin}),
we quote only the $(s \lr t)$ symmetric sum of the two exchange terms 
required by \eqn{Bhabhapermsum}.  
Again we decompose the answer according to $N_{\! f}$,
$$
\sum_{\rm spins} \biggl\{
\Re \Bigl[ \cm_4^{(2){\rm fin}} \cmt_4^{(0)\,\dagger} \Bigr]
+ U \biggl[ \Re \Bigl[ \cm_4^{(2){\rm fin}} \cmt_4^{(0)\,\dagger} 
          \Bigr] 
    \biggr]\biggr\} \ =\ 
 8 \Bigl[ \Ft^{(0)} + N_{\! f} \, \Ft^{(1)} 
         + N_{\! f}^2 \, \Ft^{(2)} \Bigr]\,. 
\equn\label{BhabhaTwoloopRemainder}
$$

In the $s$-channel, the functions $\Ft^{(i)}$ are given by
$$
\Ft^{(i)}\ =\ \Ft_s^{(i)} ,
\equn\label{FtildefromFtildes}
$$
where
$$
\eqalign{
 \Ft_s^{(0)}\ 
 &=\ - 2 {y^2\over x^2} Y^2
- 2 x^2 \Bigl( (X-Y)^2 + \pi^2 \Bigr)
\cr & \hskip0.5cm
+ {y^2\over x} 
  \Biggl[ - 4 \Bigl( \li4(-x/y) - \li4(-y) + X \li3(-y) \Bigr)
          + 2 (4 Y - 3 X - 1) \li3(-x) 
\cr & \hskip1.5cm
          + 4 \Bigl( X^2 - 2 X Y + {1\over2} X + \pi^2 \Bigr) \li2(-x)
          + {1\over8} X^4 + {4\over3} X^3 Y - 4 X^2 Y^2 
          + {2\over3} X Y^3 - {1\over6} Y^4
\cr & \hskip1.5cm
          - {23\over12} X^3 + {3\over2} X^2 Y + 9 X Y^2 - 6 Y^3
            - 5 X^2 - 21 X Y + 23 Y^2 + {93\over4} (X - 2 Y)
\cr & \hskip1.5cm
            + {\pi^2\over6} \bigl( - 17 X^2 + 32 X Y - 18 Y^2 
                                   - 17 X - 26 Y \bigr)
\cr & \hskip1.5cm
            - 2 \zeta_3  (3 X - 8 Y)   
            + 15 \zeta_4 - 38 \zeta_3 
            + 47 {\pi^2\over6} + {511\over8} \Biggr] 
\cr & \hskip0.5cm
+ {y(1-x) \over x} 
  \Biggl[ - 10 \li4(-x) + 6 X \li3(-x)
          - \biggl( X^2 + 2 {\pi^2\over3} \biggr) \li2(-x)
            + {1\over24} X^4 - {13\over12} X^3 
\cr & \hskip2.5cm
            + {\pi^2\over3} X^2  
            - {5\over2} ((X-Y)^2 + \pi^2 ) + {1\over2} Y^2  
            + \biggl( - 6 \zeta_3 + {5\over2} \pi^2 + 12 \biggr) X 
            + 20 \zeta_4 \Biggr]
\cr & \hskip0.5cm
+ 16 \Bigl( \li4(-x/y) - \li4(-y) + X \li3(-y) \Bigr)
  + 4 (3 X - 2 Y - 2) \li3(-x) 
\cr & \hskip0.5cm
  - 4 ( X^2 - 2 X Y - 2 X + \pi^2 ) \li2(-x) 
  - {5\over12} X^4 - {4\over3} X^3 Y + 8 X^2 Y^2 
  - {8\over3} X Y^3 + {2\over3} Y^4 
\cr & \hskip0.5cm
  + {5\over6} X^3 + X^2 Y 
  + 18 X^2 - 4 X Y + 2 Y^2 
  + {\pi^2\over3} ( 11 X^2 - 20 X Y + 4 Y^2 + 9 X )
\cr & \hskip0.5cm
  - 4 \zeta_3   (3 X - 2 Y)
  + 88 \zeta_4 + 8 \zeta_3 + 2 \pi^2 \ , \cr}
\equn\label{Ftildes0}
$$
$$
\eqalign{
 \Ft_s^{(1)}\ 
&=\ {1\over9} \Biggl\{ 
   {y^2 \over x} \biggl[ 
    36 \biggl( \li3(-x) - X   \li2(-x)
             - {1\over6} (X-4) X \ln(\mu^2/s)
\cr & \hskip2.5cm
             + {1\over3} (Y-3) Y \Bigl( \ln(\mu^2/s) + \ln(\mu^2/(-t)) \Bigr)
       \biggr)
           + 2 X^3 - 24 \Bigl( X +\ln(\mu^2/(-t)) \Bigr) X Y 
\cr & \hskip1.8cm
           + 12 X Y^2 - 8 Y^3
           - 19 X^2 - 58 (X-Y) Y
           + \pi^2 \Bigl( 18 X - 28 Y + 12 \ln(\mu^2/(-t)) \Bigr)
\cr & \hskip1.8cm
           + X - 66 Y + 87 \Bigl( \ln(\mu^2/s) + \ln(\mu^2/(-t)) \Bigr)
           + 58 \zeta_3 + 44 \pi^2 + {2740\over9} \biggr]
\cr & \hskip0.5cm
   - 2 x \biggl[ \biggl( 3  \Bigl( \ln(\mu^2/s) + \ln(\mu^2/(-t)) \Bigr) 
                       + 16 \biggr) X^2 
               - 6 \pi^2 X \biggr]
\cr & \hskip0.5cm
   - 4 y \biggl[ X^3  + \Bigl( 3   \ln(\mu^2/(-t)) + 8 \Bigr) X^2 
           + \biggl( 3 \Bigl( \ln(\mu^2/s) + \ln(\mu^2/(-t)) \Bigr) 
                - 2 \pi^2 + 16 \biggr) X 
            - 3 \pi^2 \biggr] \Biggr\} \, , 
\cr}
\equn\label{Ftildes1}
$$
$$
\hskip-2.5cm
 \Ft_s^{(2)}\ =\ 
{4\over9} {y^2 \over x} \biggl[ 
    \ln^2(\mu^2/s) + \ln^2(\mu^2/(-t))
      + {10\over3} \Bigl( \ln(\mu^2/s) + \ln(\mu^2/(-t)) \Bigr) 
      - \pi^2 + {50\over9} \biggr] \, ,
\equn\label{Ftildes2}
$$
and $x$, $y$, $X$, $Y$ are defined in \eqn{sChannelxyXYdef}.

In the $t$-channel, the functions $\Ft^{(i)}$ are given by the 
action of the symmetry $U$ of \eqn{Udef} on the $s$-channel results,
$$
\Ft^{(i)}\ =\ U \Bigl[ \Ft_s^{(i)} \Bigr] \,.
\equn\label{FtildetfromFtildes}
$$

In the $u$-channel, the functions $\Ft^{(i)}$ are given by
$$
\Ft^{(i)}\ =\ \Ft_u^{(i)} + U \Bigl[ \Ft_u^{(i)} \Bigr] \ ,
\equn\label{FtildefromFtildeu}
$$
where
$$
\eqalign{
 \Ft_u^{(0)}\ &=\ 
- {2 \over x^2} V^2
\cr & \hskip0.5cm
+ {y^2 \over x} 
    \Biggl[ - 4 \Bigl( \li4(-v) + W \li3(-v) \Bigr)
            + 6 V \Bigl( \li3(-v) + \li3(-w) \Bigr)
            + 2 \li3(-v) 
\cr & \hskip1.5cm
            + \biggl( - {5\over2} V^2 - 5 V W + {3\over2} W^2 - V + W
              + 11 {\pi^2\over2} \biggr) \li2(-v) 
\cr & \hskip1.5cm
            + \biggl( {1\over8} V^3 - {1\over3} V^2 W - {13\over8} V W^2
                    - {23\over12} V^2 + {21\over4} V W
                    - 5 V + {31\over2} W - 12 \zeta_3 + {93\over4} 
\cr & \hskip2.5cm
                + {\pi^2\over6} 
                   \Bigl( - 20 V + {45\over2} W - {93\over2} \Bigr) 
              \biggr) V
            + {47\over8} \zeta_4 - 20 \zeta_3 
            - 109 {\pi^2\over12} + {511\over16} \Biggr]
\cr & \hskip0.5cm
+ {y(1-x) \over x} \Biggl[  5 \li4(-v/w) 
              + 6 V \Bigl( \li3(-v) + \li3(-w) \Bigr) 
\cr & \hskip2.5cm
              - {1\over2}  \biggl( (3 V+W-2) (V-W) 
                                 + 5 {\pi^2\over3} \biggr) \li2(-v)
\cr & \hskip2.5cm
              + \biggl( - {1\over4} V^3 + {1\over2} V^2 W 
                        + {13\over12} V^2 - {9\over4} V W
                        + {5\over2} V - 12 + {\pi^2\over12} (V + 7) 
                \biggr)   V \Biggr]
\cr & \hskip0.5cm
+ 16 \Bigl( \li4(-v) + W   \li3(-v) \Bigr)
- 12 V \Bigl( \li3(-v) + \li3(-w) \Bigr)
+ 8 \li3(-v) 
\cr & \hskip0.5cm
+ \biggl( (3 V+W-4) (V-W) + 5 {\pi^2\over3} \biggr)   \li2(-v)
+ \biggl( - {5\over12} V^3 + 2 V^2 W - {9\over4} V W^2
          + {5\over6} V^2 + {1\over2} V W 
\cr & \hskip1.5cm
          + 18 V - 16 W
          + {\pi^2\over6} ( - 4 V + 9 W - 7 ) \biggr) V
- {253\over4} \zeta_4 + 8 \pi^2 \, , \cr}
\equn\label{Ftildeu0}
$$
$$
\eqalign{
 \Ft_u^{(1)}\ 
&=\ {1\over9} \Biggl\{ 
  {y^2 \over x} \biggl[ - 36 \Bigl( \li3(-v) - V   \li2(-v) \Bigr) 
              + 87 \ln(\mu^2/(-s)) + 47 \zeta_3 + {1370\over9} \biggr]
\cr & \hskip0.5cm
  - x \biggl[ 6 \Bigl( (V-W)^2 + \pi^2 + 2 (V-W) \Bigr) \ln(\mu^2/(-s)) 
\cr & \hskip1.5cm
        + (13 V - 31 W + 10 \pi^2 + 65 ) (V-W) + 16 \pi^2 \biggr]
\cr & \hskip0.5cm
  - 2 y \biggl[ - 2 V^3 + 18 V^2 W + 12 V W \ln(\mu^2/(-s))
              + 9 V^2 + 20 V W + 18 (V+W) \ln(\mu^2/(-s))
\cr & \hskip1.5cm
              + 33 V 
              + {\pi^2\over2} \Bigl( 4 V - 12 \ln(\mu^2/(-s)) - 29 \Bigr)
        \biggr] \Biggr\}\, , \cr}
\equn\label{Ftildeu1}
$$
$$
\hskip-7cm
 \Ft_u^{(2)}\ =\ 
{4\over9} {y^2 \over x} \biggl[ 
     \ln^2(\mu^2/(-s)) + {10\over3} \ln(\mu^2/(-s)) 
   + {25\over9} \biggr] \, .
\equn\label{Ftildeu2}
$$
Here $x$, $y$ are defined in \eqn{sChannelxyXYdef}, 
whereas $v$, $w$, $V$, $W$ are defined in \eqn{uChannelvwVWdef}.

\subsection{Checks on the Result}

We performed several checks on our calculation.  
The calculation was performed with the computer algebra programs
Maple, Mathematica, and FORM.  To check the code, large parts of the 
calculation were performed independently with alternative programs 
written in different languages.  Various checks were applied to the
integral reduction procedures described in 
refs.~\cite{PBScalar,NPBScalar,PBReduction,NPBReduction,AGOOne,AGOTwo}
and our implementation of them.  For example, we reproduced the 
double box ultraviolet divergences in $d=8$ and $d=10$ reported in
ref.~\cite{BDDPR}, and several other previously calculated double box
integrals~\cite{AllplusTwo}. 
An additional check on the non-planar tensor integrals is that
unphysical $1/(t-u)$ poles occur in the representation of these integrals
in terms of the master integral basis we used~\cite{NPBReduction};
however, in the series expansion in $\e$ such poles drop out after
delicate cancellations between the various terms.  

We checked the gauge invariance of the scattering amplitude by
explicitly calculating the Feynman diagrams in a general $\xi$
gauge and observing that the gauge dependence drops out in the final
result.  This provides a non-trivial check of the diagrams and
parts of the integral reduction procedure.

A strong check on the final result is provided by the matching of the 
IR divergence structure of the two-loop scattering amplitude with 
Catani's formula~(\ref{TwoloopCatani}), as discussed in 
sect.~\ref{DivergenceSubsection}.  A given integral will contribute to 
both infrared divergences and to finite terms.  Thus
a check of the divergent terms provides an indirect check 
that the finite terms have been correctly assembled.

Finally, we observed for small scattering angles a suppression
of the leading logarithms, $\ell \equiv \ln(\theta^2/4)$, e.g. in the
limit $s\to0$ in the $t$-channel for process~(\ref{eemumuKin}).  
In other small-angle limits (those not enhanced by the photon propagator
pole) the leading power-law behavior is of course less singular, but 
it is dressed by large logarithms of the type $\ell^4$ and $N_{\! f} \ell^3$.
But in the $t$-channel $s\to0$ limit it cancels down to $\ell^2$ 
and $N_{\! f} \ell$.  This behavior is in accord with a generalized eikonal
representation for small-angle scattering~\cite{AFKLMT}.

\section{Conclusions}

In this paper we presented the two-loop QED corrections to $e^+ e^-
\rightarrow \mu^+ \mu^-$ and to Bhabha scattering.  We presented the
results in terms of two-loop amplitudes interfered with tree
amplitudes and summed over spins in the context of conventional
dimensional regularization.  In these results we have set the small
electron and muon masses to vanish.  (This is an excellent
approximation for the highest energy current and future
electron-positron colliders.)

The two-loop amplitudes presented in this paper are infrared divergent.
To make use of them in a Monte Carlo program for the NNLO terms in the
cross section, they must be combined with lower-loop matrix elements
including photon emission, which should be computed using conventional
dimensional regularization, at least in the singular regions of phase
space.  In particular, the pieces that need to be computed (for the Bhabha
case) are
\begin{itemize}
\item the $e^+e^-e^+e^-$ one-loop amplitude interfered with itself.
\item the $e^+e^-e^+e^-\gamma$ one-loop amplitude interfered with
a five-point tree amplitude, and 
\item the $e^+e^-e^+e^-\gamma\gamma$ tree-level squared matrix element, 
\end{itemize}

The interference of the dimensionally regularized one-loop four-point 
amplitude with itself does not appear to be in the literature.  Nevertheless, 
it should be relatively straightforward to obtain, given that it involves 
only one-loop amplitudes with four-point kinematics.  The required
integrals are given to sufficiently high order in $\e$ in 
\eqn{OneLoopIntegralExp}.

The QED one-loop five-point amplitude interfered with the five-point tree 
is a rather involved object to compute from scratch.  However, 
the closely related one-loop helicity amplitudes for one photon and 
two quark pairs are known~\cite{KunsztFive,OneLoopPhoton,DKMPhoton},
and it is a relatively simple matter to modify the color factors to 
obtain the corresponding QED amplitudes. 
The one-loop helicity amplitudes are in the 't Hooft-Veltman scheme.
They can be converted to conventional dimensional
regularization by altering the tree amplitude appearing in the
coefficient of their singular terms~\cite{HVtoCDR}.
Thus the $e^+e^-\mu^+\mu^-\gamma$ and $e^+e^-e^+e^-\gamma$ one-loop 
amplitudes may be extracted from the known literature through 
$\Ord(\eps^0)$.

Because of the $1/\e^2$ infrared divergences that are encountered
in the phase-space integral, in regions where the photon is soft or 
collinear, one might seem to require the one-loop five-point amplitude 
through $\Ord(\e^2)$.  However, this is not necessary~\cite{LoopSplit}. 
Instead, one can replace the five-point amplitudes in singular phase-space
regions by a combination of four-point amplitudes 
(which are given in this paper to the required order in the dimensional 
regularization parameter) and splitting amplitudes~\cite{Splitting,QQGGG}.  
The one-loop splitting amplitudes for QCD are enumerated to the
required order in refs.~\cite{LoopSplit}; the case of QED follows as
usual by an appropriate conversion of color factors. 

The tree-level helicity amplitudes for $e^+e^-\mu^+\mu^-\gamma\gamma$ 
and $e^+e^-e^+e^-\gamma\gamma$ have been known for a 
while~\cite{QEDeemumugg}.  (They also can be converted from the
four-quark two photon amplitudes in ref.~\cite{DKMPhoton}, for example.)  
In infrared-divergent regions of phase space
one must include higher order in $\eps$ contributions from the matrix
elements.  Systematic discussion of these regions, where
two particles can be soft or three collinear, has been presented in
refs.~\cite{ThreeSplit} for the case of QCD. Once again the results 
for QED can be obtained by a conversion of the color factors.

Even with all of these matrix element ingredients assembled, it is a 
nontrivial task to devise a numerically stable method for carrying out
the singular phase-space integrations.  Nevertheless, this task is very
analogous to that required to obtain QCD jet predictions at
next-to-next-to-leading order, so it is likely that it will be attacked soon.

Besides the obvious application of the present paper to refined
theoretical predictions for Bhabha scattering and for
electron-positron annihilation into muons, it also serves as a further
test of methods that can be applied to analogous QCD processes.  
We are confident that many more multi-particle two-loop amplitudes
will be calculated before long.



\begin{thebibliography}{99}

\bibitem{Catani}
S.~Catani,
Phys.\ Lett.\  {\bf B427}, 161 (1998) [hep-ph/9802439].

\bibitem{AFKLMT}
A.B.~Arbuzov, V.S.~Fadin, E.A.~Kuraev, L.N.~Lipatov, N.P.~Merenkov 
and L.~Trentadue,
Nucl.\ Phys.\  {\bf B485}, 457 (1997)
[hep-ph/9512344].

\bibitem{SABSRef}
G.~Montagna, M.~Moretti, O.~Nicrosini, A.~Pallavicini and F.~Piccinini,
Nucl.\ Phys.\  {\bf B547}, 39 (1999)
[hep-ph/9811436];\\
B.~F.~Ward, S.~Jadach, M.~Melles and S.~A.~Yost,
Phys.\ Lett.\  {\bf B450}, 262 (1999)
[hep-ph/9811245].

\bibitem{MesonFact}
C.M.~Carloni Calame, C.~Lunardini, G.~Montagna, O.~Nicrosini and F.~Piccinini,
Nucl.\ Phys.\  {\bf B584}, 459 (2000)
[hep-ph/0003268].

\bibitem{Spectrum}
M.N.~Frary and D.J.~Miller,
in {\it Proceedings, $e^+e^-$ Collisions at 500-GeV, pt. A 
(Munich/Annecy/Hamburg 1991)}, report DESY 92-123A, p, 379;\\
%
N.~Toomi, J.~Fujimoto, S.~Kawabata, Y.~Kurihara and T.~Watanabe,
Phys.\ Lett.\  {\bf B429}, 162 (1998).

\bibitem{mceg}
Reports of the Working Group on Precision Calculations for the $Z$
Resonance, CERN yellow report 95-03 (1995), part 3.

\bibitem{FourLoopPrevious}
S.G.~Gorishnii, A.L.~Kataev and S.A.~Larin,
Phys.\ Lett.\ {\bf B259}, 144 (1991);\\
L.R.~Surguladze and M.A.~Samuel,
Phys.\ Rev.\ Lett.\ {\bf 66}, 560 (1991),
err. {\it ibid.} {\bf 66}, 2416 (1991);\\
T.~van Ritbergen, J.A.~Vermaseren and S.A.~Larin,
Phys.\ Lett.\ {\bf B400}, 379 (1997)
[hep-ph/9701390].

\bibitem{BRY}
Z.~Bern, J.S.~Rozowsky and B.~Yan,
Phys.\ Lett.\ {\bf B401}, 273 (1997)
[hep-ph/9702424].

\bibitem{BDDPR}
Z.~Bern, L.~Dixon, D.C.~Dunbar, M.~Perelstein and J.S.~Rozowsky,
Nucl.\ Phys.\  {\bf B530}, 401 (1998)
[hep-th/9802162].

\bibitem{AllplusTwo}
Z.~Bern, L.~Dixon and D.A.~Kosower,
JHEP {\bf 0001}, 027 (2000)
[hep-ph/0001001].

\bibitem{GloverTY}
E.W.N.~Glover and M.E.~Tejeda-Yeomans,
hep-ph/0010031.

\bibitem{PBScalar}
V.A.~Smirnov,
Phys.\ Lett.\  {\bf B460}, 397 (1999)
[hep-ph/9905323].

\bibitem{NPBScalar}
J.B.~Tausk,
Phys.\ Lett.\  {\bf B469}, 225 (1999)
[hep-ph/9909506].

\bibitem{PBReduction}
V.A.~Smirnov and O.L.~Veretin,
Nucl.\ Phys.\  {\bf B566}, 469 (2000)
[hep-ph/9907385].

\bibitem{NPBReduction}
C.~Anastasiou, T.~Gehrmann, C.~Oleari, E.~Remiddi and J.B.~Tausk,
Nucl.\ Phys.\  {\bf B580}, 577 (2000)
[hep-ph/0003261].

\bibitem{GRReduction}
T.~Gehrmann and E.~Remiddi,
Nucl.\ Phys.\  {\bf B580}, 485 (2000)
[hep-ph/9912329].

\bibitem{ReductionTalks}
T.~Gehrmann and E.~Remiddi,
Nucl.\ Phys.\ Proc.\ Suppl.\  {\bf 89}, 251 (2000)
[hep-ph/0005232];\\
C.~Anastasiou, J.~B.~Tausk and M.~E.~Tejeda-Yeomans,
Nucl.\ Phys.\ Proc.\ Suppl.\  {\bf 89}, 262 (2000)
[hep-ph/0005328].

\bibitem{AGOOne}
C.~Anastasiou, E.W.N.~Glover and C.~Oleari,
Nucl.\ Phys.\  {\bf B565}, 445 (2000)
[hep-ph/9907523].

\bibitem{AGOTwo}
C.~Anastasiou, E.W.N.~Glover and C.~Oleari,
Nucl.\ Phys.\  {\bf B575}, 416 (2000)
[hep-ph/9912251].
%

\bibitem{TwoloopOneMassIntegrals}
V.A.~Smirnov,
hep-ph/0007032;\\
%
T.~Gehrmann and E.~Remiddi,
hep-ph/0008287.

\bibitem{LoopSplit}
Z.~Bern and G.~Chalmers,
Nucl.\ Phys.\  {\bf B447}, 465 (1995)
[hep-ph/9503236];\\
Z.~Bern, V.~Del Duca and C.R.~Schmidt,
Phys.\ Lett.\ {\bf B445}, 168 (1998)
[hep-ph/9810409];\\
D.A.~Kosower and P.~Uwer,
Nucl.\ Phys.\  {\bf B563}, 477 (1999)
[hep-ph/9903515];\\
Z.~Bern, V.~Del Duca, W.B.~Kilgore and C.R.~Schmidt,
Phys.\ Rev.\ {\bf D60}, 116001 (1999)
[hep-ph/9903516];\\
S.~Catani and M.~Grazzini,
hep-ph/0007142.

\bibitem{ThreeSplit}
J.M.~Campbell and E.W.N.~Glover,
Nucl.\ Phys.\ {\bf B527}, 264 (1998)
[hep-ph/9710255];\\
S.~Catani and M.~Grazzini,
Phys.\ Lett.\ {\bf B446}, 143 (1999)
[hep-ph/9810389];
Nucl.\ Phys.\  {\bf B570}, 287 (2000)
[hep-ph/9908523].

\bibitem{GDRGlover}
A.~Gehrmann-De Ridder and E.W.N.~Glover,
Nucl.\ Phys.\  {\bf B517}, 269 (1998)
[hep-ph/9707224].

\bibitem{APSplitting}
W.L.~van Neerven and A.~Vogt,
Phys.\ Lett.\  {\bf B490}, 111 (2000)
[hep-ph/0007362];\\
A.~Retey and J.A.M.~Vermaseren,
hep-ph/0007294.

\bibitem{LeeNauenberg}
T.~Kinoshita,
J.\ Math.\ Phys.\  {\bf 3}, 650 (1962);\\
%
T.D.~Lee and M.~Nauenberg,
Phys.\ Rev.\  {\bf 133}, B1549 (1964);\\
J.~Collins, D.~Soper and G.~Sterman, in {\it Perturbative QCD}, edited by
A.H. Mueller (World Scientific, Singapore, 1989), and references therein.

\bibitem{QuarkFormFactor}
R.J.~Gonsalves,
Phys.\ Rev.\  {\bf D28}, 1542 (1983); \\
G.~Kramer and B.~Lampe,
Z.\ Phys.\  {\bf C34}, 497 (1987), 
err. {\it ibid.} {\bf C42}, 504 (1989);\\
T.~Matsuura and W.L.~van Neerven,
Z.\ Phys.\  {\bf C38}, 623 (1988);\\
T.~Matsuura, S.C.~van der Marck and W.L.~van Neerven,
Nucl.\ Phys.\  {\bf B319}, 570 (1989).

\bibitem{HarlanderggH}
R.V.~Harlander,
hep-ph/0007289.

\bibitem{HV}
G.~'t Hooft and M.~Veltman,
Nucl.\ Phys.\ {\bf B44}, 189 (1972).

\bibitem{ElectronSF}
V.N.~Baier, V.S.~Fadin and V.A.~Khoze,
Nucl.\ Phys.\  {\bf B65}, 381 (1973);\\
G.~Montagna, F.~Piccinini and O.~Nicrosini,
Phys.\ Rev.\  {\bf D48}, 1021 (1993).

\bibitem{NielsenRef}
See e.g. K.S.~K\"olbig,
SIAM J.\ Math.\ Anal.\  {\bf 17}, 1232 (1986).

\bibitem{Lewin}
L.~Lewin, {\it Dilogarithms and Associated Functions}
(Macdonald, 1958).

\bibitem{BeenakkerBhabha}
W.~Beenakker, F.A.~Berends and S.C.~van der Marck,
Nucl.\ Phys.\  {\bf B349}, 323 (1991).

\bibitem{KunsztFive}
Z.~Kunszt, A.~Signer and Z.~Tr\'ocs\'anyi,
Phys.\ Lett.\  {\bf B336}, 529 (1994)
[hep-ph/9405386].

\bibitem{OneLoopPhoton}
A.~Signer,
Phys.\ Lett.\  {\bf B357}, 204 (1995)
[hep-ph/9507442].
%

\bibitem{DKMPhoton}
V.~Del Duca, W.B.~Kilgore and F.~Maltoni,
Nucl.\ Phys.\  {\bf B566}, 252 (2000)
[hep-ph/9910253].

\bibitem{HVtoCDR}
Z.~Kunszt, A.~Signer and Z.~Tr\'ocs\'anyi,
Nucl.\ Phys.\  {\bf B411}, 397 (1994)
[hep-ph/9305239];\\
S.~Catani, M.H.~Seymour and Z.~Tr\'ocs\'anyi,
Phys.\ Rev.\  {\bf D55}, 6819 (1997)
[hep-ph/9610553].

\bibitem{Splitting}
Z.~Bern, L.~Dixon, D.C.~Dunbar and D.A.~Kosower,
Nucl.\ Phys.\  {\bf B425}, 217 (1994)
[hep-ph/9403226].

\bibitem{QQGGG}
Z.~Bern, L.~Dixon and D.A.~Kosower,
Nucl.\ Phys.\  {\bf B437}, 259 (1995)
[hep-ph/9409393].

\bibitem{QEDeemumugg}
F.~A.~Berends, P.~De Causmaecker, R.~Gastmans, R.~Kleiss, W.~Troost and 
T.T.~Wu, 
Nucl.\ Phys.\  {\bf B264}, 265 (1986);\\
J.~F.~Gunion and Z.~Kunszt,
Phys.\ Lett.\  {\bf B176}, 477 (1986).

\end{thebibliography}
\end{document}